\documentstyle[prl,aps,psfig,multicol]{revtex}
\draft
\newcommand{\be}{\begin{equation}}
\newcommand{\ee}{\end{equation}}

\newcommand{\bary}{\begin{eqnarray}}
\newcommand{\eary}{\end{eqnarray}}
\begin{document}
\title{Neutrino helicity flip by Cerenkov emission and absorption of
plasmons in supernova} 
\author{    
Subhendra Mohanty$^{*}$ and Sarira Sahu$^{\dag}$}
\address{ Physical Research Laboratory,
Navrangpura, Ahmedabad - 380 009, India}

\maketitle

\begin{abstract}
We show that in a supernova core the longitudinal photon (plasmon) has a
space-like dispersion and Cerenkov absorption and emission of such
photons is kinematically allowed. If the neutrino has a non-zero magnetic
moment, then helicity flipping Cerenkov absorption of a plasmon
$\nu_L+\gamma\rightarrow\nu_R$ is the most efficient cooling mechanism of the
supernova core, and this allows us to put a restrictive bound on the
neutrino magnetic moment $\mu_{\nu}~<~0.7\times 10^{-13}\mu_B$.

\end{abstract}

\begin{multicols}{2}
It had been pointed out by Mohanty and Samal\cite{mohanty} 
that the
helicity flipping Cerenkov process $\nu_L\rightarrow\nu_R +\gamma$ or
$\gamma +\nu_L\rightarrow\nu_R$ could be an important cooling mechanism
for the supernova core. 
Comparing the neutrino emissivity by the Cerenkov process 
with observations of SN1987A, a restrictive bound  
on the neutrino magnetic moment was established. 
Subsequently it was pointed out by Raffelt\cite{raffelt} that,
this result was based on a numerical error in the calculation
of the refractive index of the SN core and using the correct numbers it was
shown that the photons in a SN core do not have a space-like 
dispersion relation, so the
Cerenkov helicity flip process would not occur. 

Here we show that the
earlier estimate of refractive index  was based on the thermodynamic formula
for susceptibility which turns out to be 
invalid for real photons or plasmons even in the
static limit. However an analysis of the dispersion relations of plasmons in
an ultra relativistic plasma\cite{bratten} 
shows that the longitudinal photon (plasmon)
has a space-like branch and therefore in such a plasma the Cerenkov 
radiation of a plasmon is kinematically allowed\cite{sarira}. 
We compute the neutrino
emissivity of SN core by the Cerenkov helicity flip emission of a plasmon 
and subsequent escape of the $\nu_R$.
We show that the observations of neutrino flux from SN1987A put a constraint
on the neutrino magnetic moment $\mu_{\nu} < 0.7\times 10^{-13}\mu_B$.

The plasmon self energy diagram in a degenerate plasma 
is given by\cite{bratten}
\be
\Pi_L(k,w)=\frac{\tilde\mu_e^2 e^2}{\pi^2}
\left ( 1 - \frac{w^2}{k^2}\right )
\left [1- \frac{w}{2 k}\ln \left(\frac{w+k}{w-k}\right )\right ]
\ee
where $\tilde\mu_e$ is the electron chemical potential which we have assumed 
to be larger than $m_e$ and $T$. The dispersion relation for the 
longitudinal propagating mode $w^2-k^2=Re~\Pi_L(k,w)$ is plotted in Figure 1.
There is a lower branch which is space like. It is true that, there is also
a large imaginary part of $\Pi_L$ when $w^2<k^2$, which means that this mode
is Landu damped. However for our 
application all we need is that there
must be an emission 
or an absorption of a plasmon accompanied be a neutrino
helicity flip. 
\vspace*{4.2in}

\begin{figure}
\includegraphics{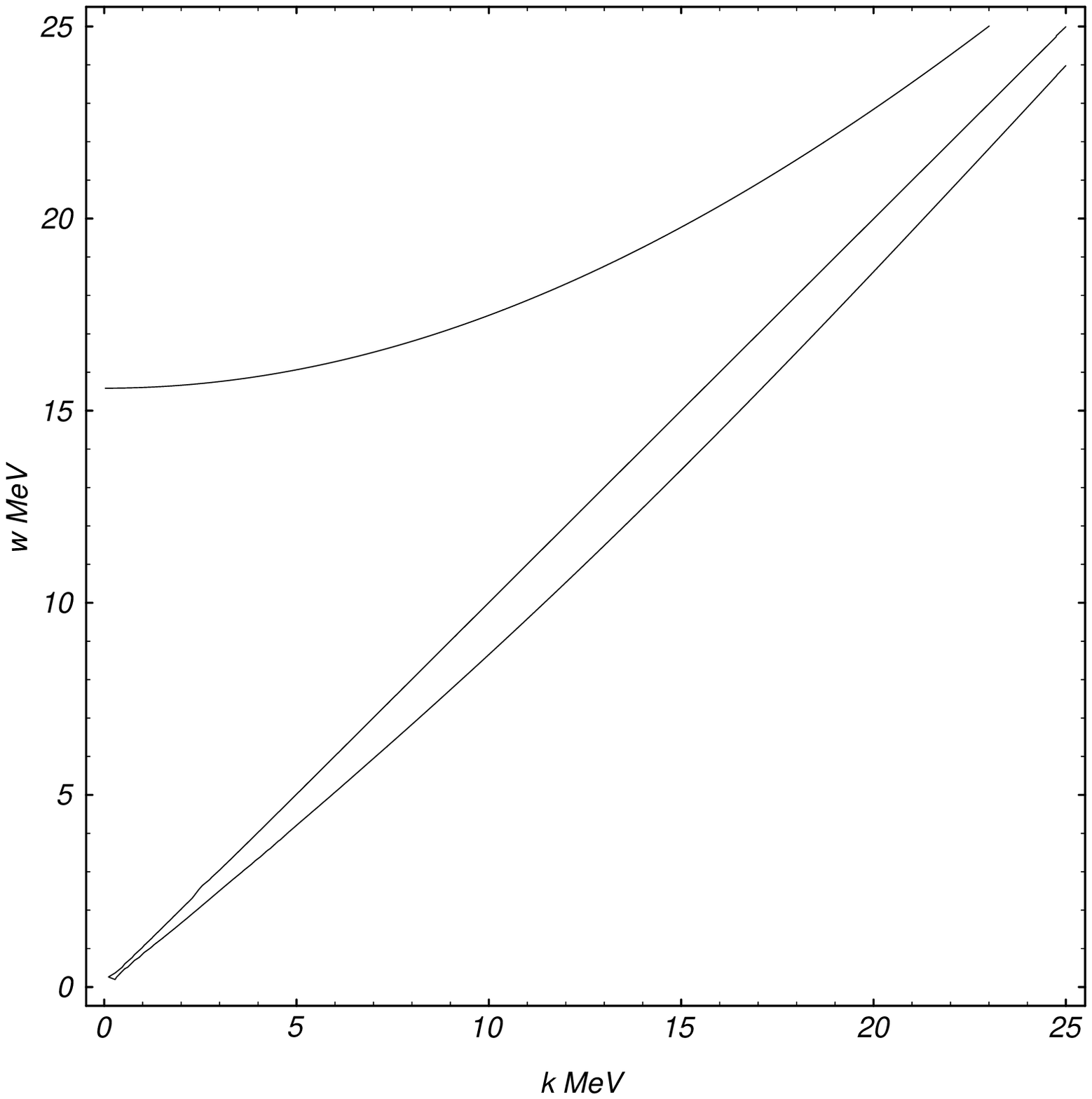}
\end{figure}
\vspace*{-2.0in}
\noindent {\it Fig. 1. Dispersion relation for the degenerate plasma}
\\
\vspace*{4.0in}
\begin{figure}
\includegraphics{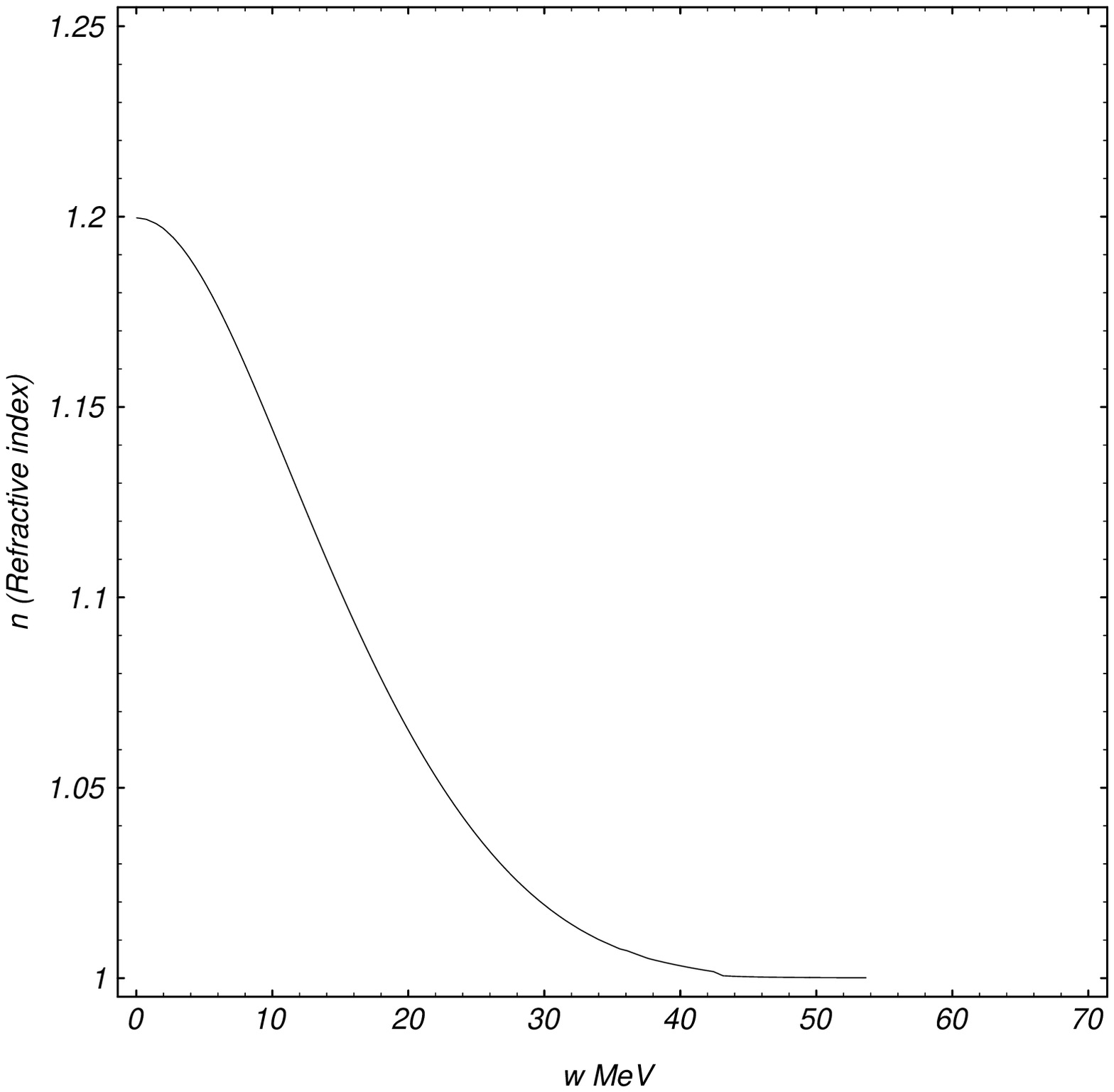}
\end{figure}
\vspace*{-1.6in}
\noindent {\it Fig. 2. 
Plasmon refractive index as a function of frequency
in a supernova core.}\\

\noindent So it is not essential for our application that plasmon need
to propagate far. Using this dispersion relation we find that in a 
supernova core there is a
range of frequencies $(0.3 - 50)~ MeV$ for which the refractive index
$n=k/w > 1$ as shown in Figure 2 and therefore plasmon emission or 
absorption by the Cerenkov
process is kinematically allowed in this range of plasmon frequencies.

The matrix element for the 
helicity flipping Cerenkov process $\nu_L(p_1)\rightarrow\nu_R(p_2) 
+\gamma (k)$ is
\be
{\cal M} = \mu_{\nu} {\bar u}(p_2)\sigma^{\mu\nu}u(p_1)
\sqrt{Z_l}\epsilon_{\mu}(k, \lambda).
\ee
The photon polarisation sum in the medium for the longitudinal part is
given by
\be
\sum_{\lambda} \epsilon^L_{\mu}\epsilon^{L*}_{\nu}
=\left (1-\frac{1}{n^2}\right )u_{\mu}u_{\nu} +
\frac{(u_{\mu}k_{\nu} + k_{\mu}u_{\nu})}{n^2 w},
\ee
where $u_{\mu}$ is the four velocity of the medium.
The rate of energy emission by this process is
\be
{\dot S}=\frac{\mu^2_{\nu}}{16\pi E^2_1}
\int (E_1-w) w^2 dw Z_l f(n) 
(w-2 E_1)^2
\ee
where $f(n)=(n^2-1)^2/n^2$ and 
the wave function renormalisation factor $Z_l$ is given by\cite{bratten}
\be
Z_l=\frac{2}{n^2 + \frac{3 m^2_{\gamma}}{w^2}-1},
\ee
where $m_{\gamma}=e^2{\tilde\mu}^2/{3 \pi^2}$ is the effective
photon mass in the medium.
The neutrino luminosity is given by
\be
Q_{\nu_R}=V\int_0^{\infty}4\pi E^2_1 dE_1~ f_{\nu}(E_1) {\dot S}(E_1),
\ee
where $f_{\nu}(E_1)$ is the distribution function of the neutrinos.
Taking the SN core volume to be
$V=4.19\times 10^{18}~ cm^3$, the electron chemical potential
$\tilde \mu_e=280$ MeV and the neutrino chemical potential 
$\tilde\mu_{\nu}=160$ MeV\cite{mohapatra} 
we obtain the neutrino luminosity to be
\be
Q_{\nu_R}=\mu^2_{\mu} \left (1.45\times 10^{61}~MeV^4 \right),
\ee
in terms of the magnetic moment of the neutrino. Assuming that the entire
energy of the core collapse is not carried away by the right handed 
neutrinos $i.e.$, $Q_{\nu_R} < 10^{52}~ergs/sec$ we obtain the upper 
bound on the neutrino magnetic moment 
$\mu_{\nu} <0.2\times10^{-11}\mu_B$.
This is comparable to the bound $(0.2 - 0.8)\times 10^{-11}\mu_B$
obtained\cite{mohapatra} from the cooling of SN1987A by the helicity flip
scattering.

At high temperature the rate of Cerenkov absorption process
$\nu_L(p_1) + \gamma (k)\rightarrow\nu_R (p_2)$ is much larger than the
Cerenkov emission. The cross section for the Cerenkov absorption is
given by
\be
\sigma (p_1, k)=\frac{\pi}{2 E_1 w}\delta((p_1+k)^2-m_{\nu}^2)|{\cal M}|^2
\ee
where 
\be
|{\cal M}|^2=\mu^2\frac{(n^2-1)^2}{n^2}Z_l w^2 (2 E_1 + w)^2.
\ee
The rate of energy carried away by $\nu_R$ is given by
\be
Q_{\nu_R}=V\int d^3p_1 d^3k f_{\nu}(p_1) f_{\gamma}(k)
\sigma (p_1, k) (E_1+w),
\ee
where $f_{\nu}(p_1)$ and $f_{\gamma}(k)$ are the distribution functions 
of the incoming $\nu_L$ and $\gamma$ respectively. The expression for 
$Q_{\nu_R}$ can be written analytically as
\bary
Q_{\nu_R} &=& V 2\pi^3 \mu^2 T^7
\int^{\infty}_0 dy\int^{x_2}_{x1} dx g(x,y) Z_l
\nonumber\\
&&
\times\left (\frac{1}{e^{y-{\tilde\mu_{\nu}/T}}+1}\right )
\left (\frac{1}{e^x-1}\right ),
\label{ceab}
\eary
where the function $g(x,y)$ is defines as
\be
g(x,y)= x^2 ~(x+y) (x+2 y)^2 (n^2-1)^2
\ee
and
$x=w/T$, $y=E_1/T$ are dimensionless quantities.
The $x$ integration is carried over the range where $n>1$ and the Cerenkov
absorption is kinematically allowed.
Taking $\tilde\mu_{\nu_L}=160$ MeV, $T=60$ MeV
inside the supernova core\cite{mohapatra} and doing the
integral in Eq.(\ref{ceab}) numerically we obtain the 
neutrino emissivity
\be
Q_{\nu_R}=\mu^2_{\nu} \left (1.1\times 10^{64} MeV^4 \right )
\label{ener}
\ee
in terms of the neutrino magnetic moment. Using the same constraint 
$Q_{\nu_R} < 10^{52}~erg/sec$ we obtain from Eq.(\ref{ener})
the upper bound on neutrino magnetic moment
\be
\mu_{\nu} ~<~0.7\times 10^{-13}\mu_B
\ee
This is two orders of magnitude lower than the earlier bounds\cite{mohapatra}.
This shows that if the neutrino has a non-zero magnetic moment
then helicity flipping of the neutrinos by the Cerenkov absorption of 
plasmons is the most efficient cooling mechanism in a supernova.

Helicity flipping process in various types of plasmas have been studied
using the quantum kinetic equation in ref\cite{raffelt1}.

\end{multicols}
\end{document}